\documentclass[12pt,preprint]{aastex}

\newcommand{\eg}{{e.g.,}\ }

\shorttitle{Ca Variations in G29-38}
\shortauthors{von Hippel \& Thompson}

\begin{document}

\title{Discovery of Photospheric Calcium Line Strength Variations in the
DAZd White Dwarf G29-38}

\author{Ted von Hippel\altaffilmark{1,2} and Susan E. Thompson\altaffilmark{3}}

\altaffiltext{1}{Department of Astronomy, University of Texas at Austin, 1
University Station C1400, Austin, TX 78712-0259, USA;
ted@astro.as.utexas.edu}
\altaffiltext{2}{Visiting Scientist, Southwest Research Institute,
1050 Walnut St., Suite 400 Boulder, CO 80302} 
\altaffiltext{3}{The Colorado College, 14 E. Cache La Poudre, Colorado 
Springs, CO 80903}

\begin{abstract}

Metals in the photospheres of white dwarfs with $T_{\rm eff}$ between
12,000 and 25,000 K should gravitationally settle out of these atmospheres
in 1--2 weeks.  Temporal variations in the line strengths of these
metals could provide a direct measurement of episodic metal accretion.
Using archival VLT and Keck spectroscopy, we find evidence that the
DAZd white dwarf G29-38 shows significant changes in its Ca II K line
strength.  At the two best-observed epochs, we find that the Ca line
equivalent width (EW) = 165 $\pm$ 4 m\AA\ (in 1996.885) and 280 $\pm$
8 m\AA\ (in 1999.653), which is an increase of 70\%.  We consider the
effect that pulsation has on the Ca EWs for this known variable star,
and find that it adds an error of $<$ 1 m\AA\ to these measurements.
Calcium line strengths at other observational epochs support variations
with timescales as short as two weeks.  These Ca EW variations indicate
that the metal accretion process in G29-38, presumably from its debris
disk, is episodic on timescales of a few weeks or less, and thus the
accretion is not dominated by Poynting-Robertson drag from an optically
thin, continuous disk, which has a timescale of $\sim1$ year.

\end{abstract}

\keywords{accretion, accretion disks --- stars: individual (G29-39) --- white dwarfs}

\section{Introduction}

The presence of Ca, Mg, Fe, or other heavy elements in the photospheres
of hydrogen-dominated (DA) white dwarfs (WDs) poses a long-standing
problem.  The high surface gravities of WDs cause their atmospheres to
become highly stratified and theory predicts that these heavy elements
will settle on timescales of $10^{-2}$ to $10^6$ years \citep{DFP92,
KW06}, depending on the mass and surface temperature of the white dwarf.
Yet most WDs are old, and almost every known DAZ, as these stars are now
called \citep{SGL83}, has been cooling as a WD for at least $10^8$ years,
and usually $10^9$ years or more.  The timescale problem is less extreme
for metals in the rarer helium-dominated (DB) white dwarfs, since the
depletion of metals due to diffusion at the bottom of the He-convection
zone is slower than at the bottom of the H-convection zone and since the
higher transparency of helium atmospheres makes lower metal abundances
more visible.

Heavy elements have been known in WD atmospheres for almost 50
years \citep[in the DB vMa 2;][]{W58}, though the first convincing
demonstration of metals in a DA was not until \cite{LWF83}.  It took
another 14 years to find the second and third DAZs \citep{KPS97, HBG97}.
The DAZ sample then began to grow with the breakthrough survey of
\cite{ZR98}, which yielded seven to nine new DAZs.  Subsequent modern
surveys \citep[\eg][]{ZKR03} have shown that $\sim$25\% of all single DAs
possess metals in their photospheres.  On the theoretical side, early work
focused on understanding how WDs might accrete sufficient material from
the interstellar medium (ISM) to maintain metals in their atmospheres.
Some researchers \citep{DFW93, KW06} concluded that ISM accretion
is the source of the photospheric metals, while other researchers
\citep{AKH93, ZR98, ZKR03, KR07} concluded that ISM accretion is
insufficient.  An alternative hypothesis, that DAZs accrete their metals
from circumstellar material, has gone from an intriguing possibility
based on just one DAZ with a debris disk \citep[G29-38,][]{ZB87}, to a
compelling scenario, especially now that five debris disk white dwarfs
are known, all of which happen to be DAZs \citep{BFJ05, KvHL05, KvHL06,
vHKK07, KR07}.  These DAZs with debris disks are now known as DAZd white
dwarfs \citep{vHKK07}.  The newly discovered debris disk WDs, the high
relative frequency of DAZs and their plausible connection to circumstellar
debris \citep[\eg][]{J03}, as well as new catalogs of DAZs \citep{ZKR03,
KRN05}, have created new opportunities to understand the origin of DAZs
and their accretion processes.

In this paper we study archival time-series spectroscopy from multiple
epochs of the luminosity variable DAZ white dwarf G29-38 (=WD2326+049).
The case for Ca line strength variations in G29-38 is complicated since
this star is a pulsating WD.  The parameters for G29-38 are $T_{\rm eff}$
= 12100, log(g) = 7.9, log(Ca/H) = $-6.8$ \citep{KRN05}, distance =
19 pc \citep{HD80}, or log(g) = 8.14 \citep{BWL95}.  The timescale for
gravitational settling in this star is 7 days according to \cite{KW06},
shorter for the higher log(g) of \cite{BWL95}, and therefore Ca line
strength variations are plausible.  After accounting for the effects of
pulsation, we find clear evidence that the Ca II K line strength varies
in this star.  Metal line strength variability is a new tool for the
study of accretion onto WDs and it may eventually help us unravel the
source(s) of metals found in DAZ atmospheres.

\section{Observations}

Both \cite{vKCW00} and \cite{TvKC07} published time-series spectroscopy
of G29-38 in order to measure the spherical degree of pulsation modes
via line shape variations of the Balmer series.  We use these spectra to
measure the Ca II K line strength present in this star.  On November 19,
1996, \cite{vKCW00} obtained 4.72 hours of continuous (no readout gaps)
12s exposures of G29-38 with the Low Resolution Imaging Spectrometer on
the Keck II Telescope.  The resolution was set by the seeing to be $\sim7$
\AA, and the average spectrum has a signal-to-noise of 1000 measured
at the continuum near 5000 \AA.  For the same purposes, on August 27,
1999, \cite{TvKC07} obtained 6.14 hours of continuous spectroscopy with
the FORS1 spectrograph on the VLT \citep[see also][]{T06}.  Each of the
604 spectra has an exposure time of 16~s and a resolution of $\sim8$
\AA.  The average spectrum has a signal-to-noise of 500 measured at
the continuum near 5000 \AA. \cite{T06} and \cite{TvKC07} performed
standard reductions on both sets of data, except that flat fields and
traditional flux calibrations were not possible for either dataset.
In both cases the average spectra are visibly smooth near 4000 \AA, and
we have determined that the lack of flat-fields has a negligible effect
on our EW measurements.  In both datasets, the Ca II 3933 \AA\ line is
clearly visible, a fact that we utilize to demonstrate EW variations
between these two observations.

\section{Calcium Equivalent Width Measurements}

In Figure 1 we present the average Ca II K line region of G29-38 at
two epochs, 1996.885 and 1999.653.  The average spectrum taken with the
Keck telescope in 1996 shows an obvious Ca II K line with EW = 165 $\pm$
4 m\AA.  The average spectrum taken with the VLT in 1999 shows a deeper
Ca line with EW = 280 $\pm$ 8 m\AA, an increase of 70\%.  Formally, these
two observations differ by 12.9 $\sigma$.  Each are measured by fitting
a Gaussian to the spectral line and the errors are those associated with
that fit.

We need to ensure that the pulsations of G29-38, which produce brightness
variations as high as 3\% for a single pulsation, are not responsible for
the changes in Ca line strength.  The pulsating modes cause changes in
surface temperature and radial velocity fields, both of which can alter
the measured absorption lines.  Since the 1996 and 1999 data sets are
time series of spectra, we measure how the EW changes over the pulsation
cycle and how this change is reduced by integrating over the extended
periods of the Keck and VLT observations.  Using these measurements,
we demonstrate (see below) that the pulsations have a negligible effect
compared to the observed 70\% change in EW of the average spectra.

We measured the equivalent width for each spectrum of both the Keck and
VLT data sets by fitting a Gaussian to the calcium line after normalizing
the spectra.  We fit an 8th-order polynomial across 20 \AA\ on each side
of the Ca II K line of the average spectrum.  We removed the curvature
induced by the broad hydrogen lines on either side of the Ca line by
dividing each spectrum by this fit to the average.  These flattened
spectra were normalized as we fit the Gaussian function by allowing the
overall flux of the spectrum to vary during the fit.  During the fit,
the area and FWHM of the Gaussian were allowed to vary.

We performed a Fourier Transform (FT) on the time-series Ca EW
measurements (Figure 2).  These FTs provide a clear indication of how
much the line strength can change during a single observing run due to
pulsation effects.  The EW FTs clearly show that the pulsations are
stronger during the Keck epoch.  The flux amplitude of the largest
observed pulsation (measured at 615s) in the VLT data is 2.7\%
\citep{TvKC07} while the largest mode (measured at 614s) in the Keck
data is 3.17\% \citep{vKCW00}.  Therefore, one would expect the Keck
observations to exhibit larger variations in Ca equivalent width.
The equivalent width amplitude of these largest flux modes are 7.1\%
and 13.8\%, with the Keck amplitude being 1.9 times larger than the
VLT amplitude.  See Table 1 for the amplitude of the EW variations
measured at the dominant modes in each data set.  Finally, we calculate
the average value of the measured equivalent widths, weighted by the
errors established from fitting the Gaussian.  The values 167 $\pm$
1.7 m\AA\ for the Keck epoch and 272 $\pm$ 1.9 m\AA\ for the VLT epoch
(an increase in 63\%) are in close agreement with the equivalent width
of the unweighted average spectrum.

The pulsations cause deviations from the average equivalent width.
Observing an incomplete cycle of a dominant pulsation can result in an
abnormally high or low measurement of the average EW.  To determine how
large this effect could be, we calculated a worst case for this star.
This worst case provides an outer bounds for our data set (useful for our
error analyses, below) and should help future observers who wish to repeat
this experiment with G29-38, possibly with long exposures that do not
resolve the pulsation modes.  We combined six modes with semi-amplitude
(mean to maximum) EW variations of 8, 12, and 15\% (2 each) and we
chose periods from 1150 to 600 seconds based on previous observations
(see Figure 3 for the specific modes).  Since the modes of G29-38 vary in
strength with time, this allows for maximal pulsations (which are larger
amplitude modes than we actually observed in the Keck and VLT data).
This approach also encompasses the possibility of observing a few large
amplitude modes along with a few moderate amplitude modes.  Figure 3
shows the possible range of EW values in percentage terms ((max EW - min
EW) / mean EW) due to only partial averaging over pulsation modes, as a
function of cumulative exposure times.  For an extremely short exposure
time, we could observe an EW difference as large as $\sim$100\% given the
semi-amplitudes of the modes introduced in this example.  After one hour
of cumulative exposure, the maximum expected EW difference due to the
pulsation modes of this star drop to 5\%.  After 3.5 hours this effect
is $<$ 1.5\%.

Since the VLT and Keck spectra are time series, we know the size of the Ca
EW variations and can better estimate the error associated with partial
averaging.  Including only the three largest pulsations of the Keck data
and averaging over a total of 4.7 hours spaced in 24s intervals (identical
to the Keck observations), we find a deviation of less than 0.3\% in
the EW.  This translates to at most an additional error of 0.5 m\AA\
and 0.8 m\AA\ for the Keck and VLT measurements, respectively.  This is
equivalent to one third of the error in the weighted average and one
tenth of the error from fitting the average spectrum.  We conclude that
partial averaging over the pulsations cannot account for the observed 70\%
variation between the measured Ca EW in 1996 at Keck and 1999 at the VLT.

In summary, we find that the EW of Ca increased by 63-70\% between
1996.885 and 1999.653.  The difference between these two estimates
is likely dominated by subtle differences in how we measure EWs, and
this range is a useful estimate of the accuracy in the measured change
in Ca EWs.  Only a negligible amount of this 63-70\% change is due to
residual, unaccounted-for pulsation effects.

Besides these two epochs with time-series spectroscopy, we were able to
derive Ca EWs for nine additional spectra of G29-38, kindly provided to
us by D. Koester. Two spectra of 30 minute duration each were observed
with the Calar Alto 2.2m Cassegrain spectrograph on September 18 and 20,
1995.  These spectra were the source of the discovery by \cite{KPS97}
that G29-38 is a DAZ.  Due to the temporal proximity of the Calar Alto
observations, we averaged together the spectra before measuring the
Ca EW.  This averaging also decreased the spectral noise and reduced the
pulsation error.  Five spectra of 30 minute duration each were observed
with Keck, one on the night of July 7, 1997, one on the night of December
10, 1998, and three on the night of August 12, 1999.  To increase the
signal-to-noise for August 12, 1999, we averaged together the spectra
from this night before measuring the Ca EW.  Two additional spectra of
5 minute duration were obtained at VLT/UVES on the nights of August 6,
2000 and September 17, 2000.  Further details on the observations and data
quality for these spectra can be found in \cite{KNC01}, \cite{KRN05},
and \cite{ZKR03}.  From these nine additional spectra we derive Ca
EWs at six additional epochs.  All of these measured Ca EWs, as well
as the two derived from the time-series spectroscopy discussed above,
are presented in Figure 4.  The error bars represent the 1 $\sigma$
quadrature combination of the Ca EW measurement uncertainty and the
expected EW variation caused by sampling over partial luminosity periods.
Figure 3 represents the maximum observed error from partial averaging
and thus we use these values as a 3$\sigma$ estimate for this error for
all epochs except the two discussed above, where we directly measure
the amplitude of the pulsation modes.

\section{The Source of Calcium Line Strength Variations}

Since G29-38 has a debris disk known to be rich in refractory elements
\citep{RKvH05}, it is natural to assume that the Ca variations we see
are due to time-variable accretion from the disk.  There is, however,
another possibility worth considering---that this object has a large
star spot that covers more or less of the observed face of G29-38 at the
different observational epochs.  Such star spots models have been proposed
before to explain variations in He line strength in three WDs with mixed
H/He atmospheres, G104-27 \citep{KHB92}, PG1210+533 \citep{BWB94}, and
HS0209+0832 \citep{HNL97}.  For a model of G29-38 with steady-state
accretion and a magnetically constrained star spot, the spot could either
be enriched in Ca with respect to the rest of the star if accretion
occurs within the spot, or it could be depleted in Ca if accretion occurs
elsewhere.  Since the variation in Ca EW is nearly a factor of two (Fig.\
4), even in a model with dramatically different Ca abundances within
and outside the spot, the spot would have to cover nearly half of a WD
hemisphere.  It seems unlikely that such a spot could persist for years,
particularly for this pulsating WD with pulsations sloshing across the
surface every few minutes.  A prediction of the star spot model is that
the Ca EW should be modulated on the timescale of the rotation period.
This effect would be particularly apparent with observations covering two
or more rotation periods.  G29-38 appears to be rotating with v sin(i) =
11--28 km s$^{-1}$ \citep{BKN05}, or a period of $\leq$ 900--2400 s or
2000--5000 s, assuming a radius consistent with log(g) = 8.14 or 7.9,
respectively.  Our VLT and Keck time-series observations cover 4.72 and
6.14 hours, so if the star spot model were correct, the rotation period
would have to be meaningfully longer than 6 hours, which is inconsistent
with the measured rotation velocity.  In summary, we consider the star
spot scenario highly unlikely, yet future observations with better
temporal sampling will be needed to convincingly rule it out.

The history of Ca EWs in G29-38 presented in Figure 4 shows that the
Ca line strength varies by at least a factor of two, that significant
variations occur on timescales as short as 15 days (August 12, 1999 to
August 27, 1999), and possibly that there is a typical Ca EW $\approx$
230 m\AA\ with periods of both higher and lower Ca line strength.
Assuming that the Ca EW variations are caused by time-variable accretion
from G29-38's debris disk, the low Ca EW in late 1996 indicates
that accretion decreased dramatically for at least one settling time
\citep[expected to be $\sim$ 7 days;][]{KW06}.  The variations in August
1999 similarly indicate an increased accretion rate with a timescale of
$\sim$ 2 settling times.  Further observations and analysis are required
for this star in order to determine the minimum timescale and range for
accretion events, as well as to observationally test the theoretical
gravitational settling time.

We note also that D. Koester kindly searched through multiple spectra of
24 DAZs from the SPY survey \citep{NCD01, KRN05} for us in order to search
for other WDs with varying Ca EWs.  He found no other convincing cases.
Perhaps a larger sample of stars or more spectra per star are needed.
In any case, further examples besides G29-38 would help clarify whether
episodic accretion is a function of other system parameters and would
help test the theoretically determined residence timescale for metals
in DAZ atmospheres.

\section{Conclusions}

We find that the Ca II K line in G29-38 varies from EW = 165 $\pm$ 4
m\AA\ in 1996.885 to EW = 280 $\pm$ 8 m\AA\ in 1999.653, or a factor
of $\sim 1.7 \times$ between these two epochs.  Fully taking into
account variations in G29-38's EW due to sampling effects over its
multi-period pulsations, we find that the EW variations are essentially
unaffected over these 4.72 hour and 6.14 hour time-series observations.
We also measure Ca EWs in this star for a further six epochs and find
convincing evidence that the Ca line changes on timescales as short as
15 days.  These Ca line-strength variations indicate that the source
of the variations, presumably metal accretion from the debris disk,
is not a steady-state process and varies on timescales of a few weeks.
Observations with greater temporal sampling are required to determine
the timescale range of G29-38's episodic accretion.

Our Ca EW measurements may already provide insight into the physics
governing accretion onto this star.  For instance, if accretion is
dominated by Poynting-Robertson (P-R) drag from the known debris disk,
then the dust accretion timescale is $4 r^2 a$ years \citep{RKvH05},
where $r$ is the particle distance in solar radii and $a$ is the particle
radius in microns.  Spitzer mid-IR spectroscopy \citep{RKvH05} indicates
sub-micron silicate particles.  With an inner debris disk edge at $\leq$
0.15 $R_{\sun}$ \citep{vHKK07}, the P-R timescale at the inner disk edge
is $\leq$ 33 days, consistent with the Ca EW variations, especially if
whatever process feeds dust to the inner disk edge is discontinuous.
This scenario argues against an optically thin (in the radial direction)
debris disk extending continuously to 1 $R_{\sun}$ or beyond.  Comparing
these timescales indicates that G29-38's debris disk is likely to be
optically thick with collisional processes or dust delivery radially
through an optically thick disk setting the accretion timescale.

Due to the short settling times of metals in their atmospheres, warm
($T_{\rm eff}$ = 12,000--25,000~K) DAZs are an excellent place to look
for Ca (or other metal) line strength variations.  Will other DAZs that
harbor debris disks show Ca line strength variations?  And if so, will
the timescale or range of these variations be correlated with any of
the debris disk properties?  Will DAZs without detectable debris disks
show Ca line strength variations?  Answers to these questions will help
us understand the accretion process operating in these systems and
possibly give us clues to the origins of white dwarf surface metals.
Additionally, the timescale of metal line strength variations as
a function of stellar effective temperature would provide the first
observational test of the theoretically determined residence timescale
for metals in DAZ atmospheres.

\acknowledgments
We thank Detlev Koester for his many excellent spectra of G29-38, and for
considerable consultation on WD models and this paper.  We thank Marten
van Kerkwijk for his reductions of the Keck data and Mike Montgomery
for helpful discussions.  We thank our referee, Ralf Napiwotzki, for his
insights and for comments that improved the presentation.  TvH gratefully
acknowledges support as a visiting scientist from the Southwest Research
Institute.  This material is based upon work supported by the National
Aeronautics and Space Administration under Grant No.\ NAG5-13070 issued
through the Office of Space Science.

\bibliography{apjmnemonic,references}
\bibliographystyle{apj}

\clearpage

\begin{deluxetable}{lcc}
\tablewidth{0pt}
\tablecaption{EW Variations for Different Periods in the Keck and VLT Data.}
\tablehead{
\colhead{period} & \colhead{Amp (m\AA)} & \colhead{Amp (\%)} \\
\phantom{123} (1) & \phantom{1} (2) & (3) }
\startdata
Keck (1996.885) &&\\
614.25  & 23 $\pm$ 3 & 13.81 $\pm$ 1.8 \\
817.66  & 18 $\pm$ 3 & 10.03 $\pm$ 1.8 \\
653.08  &  5 $\pm$ 3 &  2.95 $\pm$ 1.8 \\
775.97  & 10 $\pm$ 3 &  5.24 $\pm$ 1.8 \\
VLT (1999.653) &&\\
615.371 & 20 $\pm$ 4 &  7.1  $\pm$ 1.4 \\
810.759 & 20 $\pm$ 4 &  7.4  $\pm$ 1.4 \\
835.283 & 17 $\pm$ 4 &  6.2  $\pm$ 1.4 \\
353.469 &  1 $\pm$ 4 &  0.4  $\pm$ 1.4 \\
\enddata
\tablerefs{We fit the series of EW measurements with periods found by
\cite{vKCW00} and by \cite{TvKC07} for the Keck and VLT observations,
respectively.  We present the amplitude of the changing EWs in
milli-{\AA}ngstroms and in percentages.  The errors represent the formal
errors from the fits.} 
\end{deluxetable}

\clearpage

\begin{figure}[!t]
\plotone{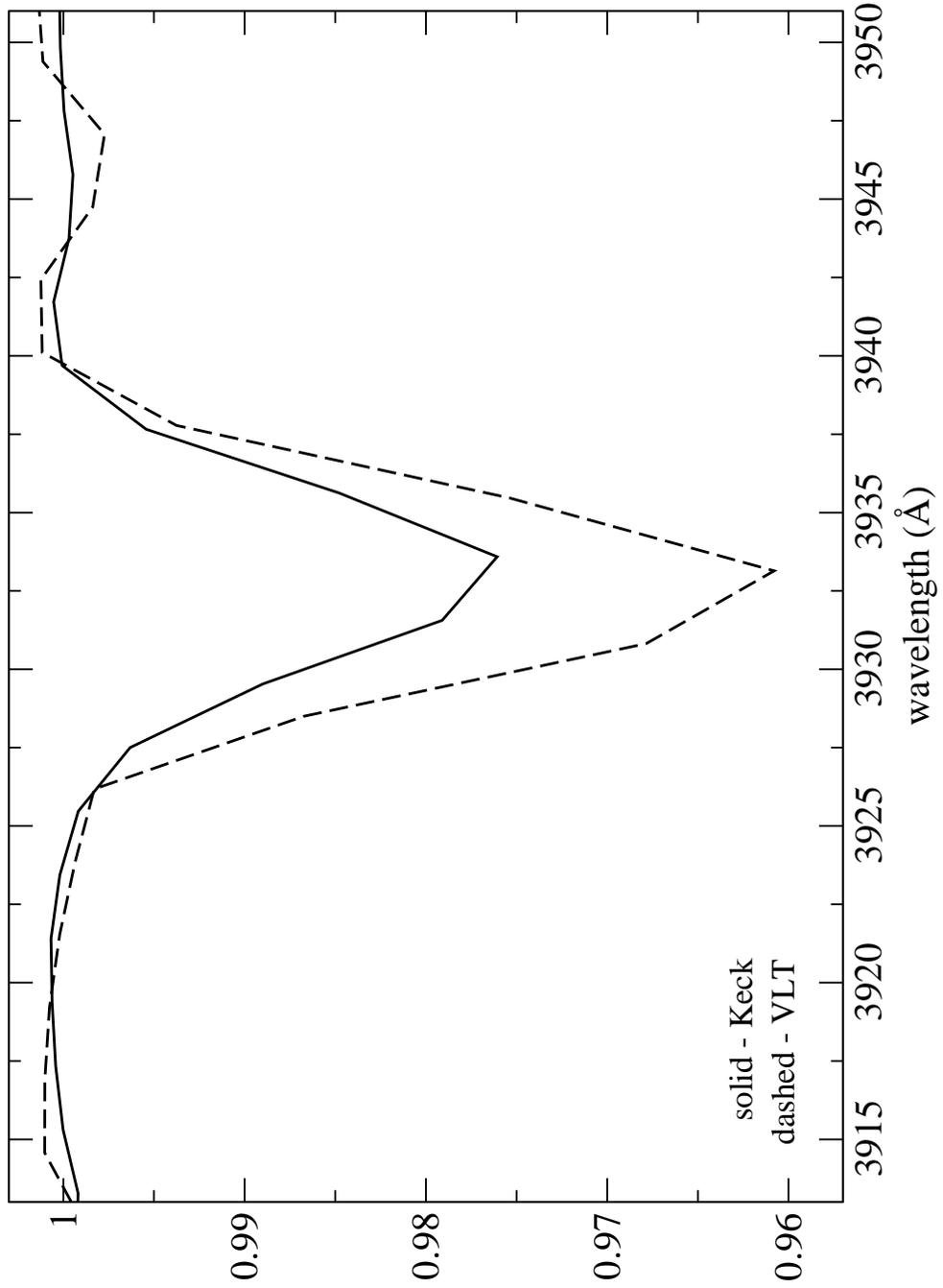}	
\figcaption{The Ca line region in G29-38 at two different epochs,
1996.885 and 1999.653.  The data are the averaged, normalized spectra.}
\end{figure}

\begin{figure}[!t]
\plotone{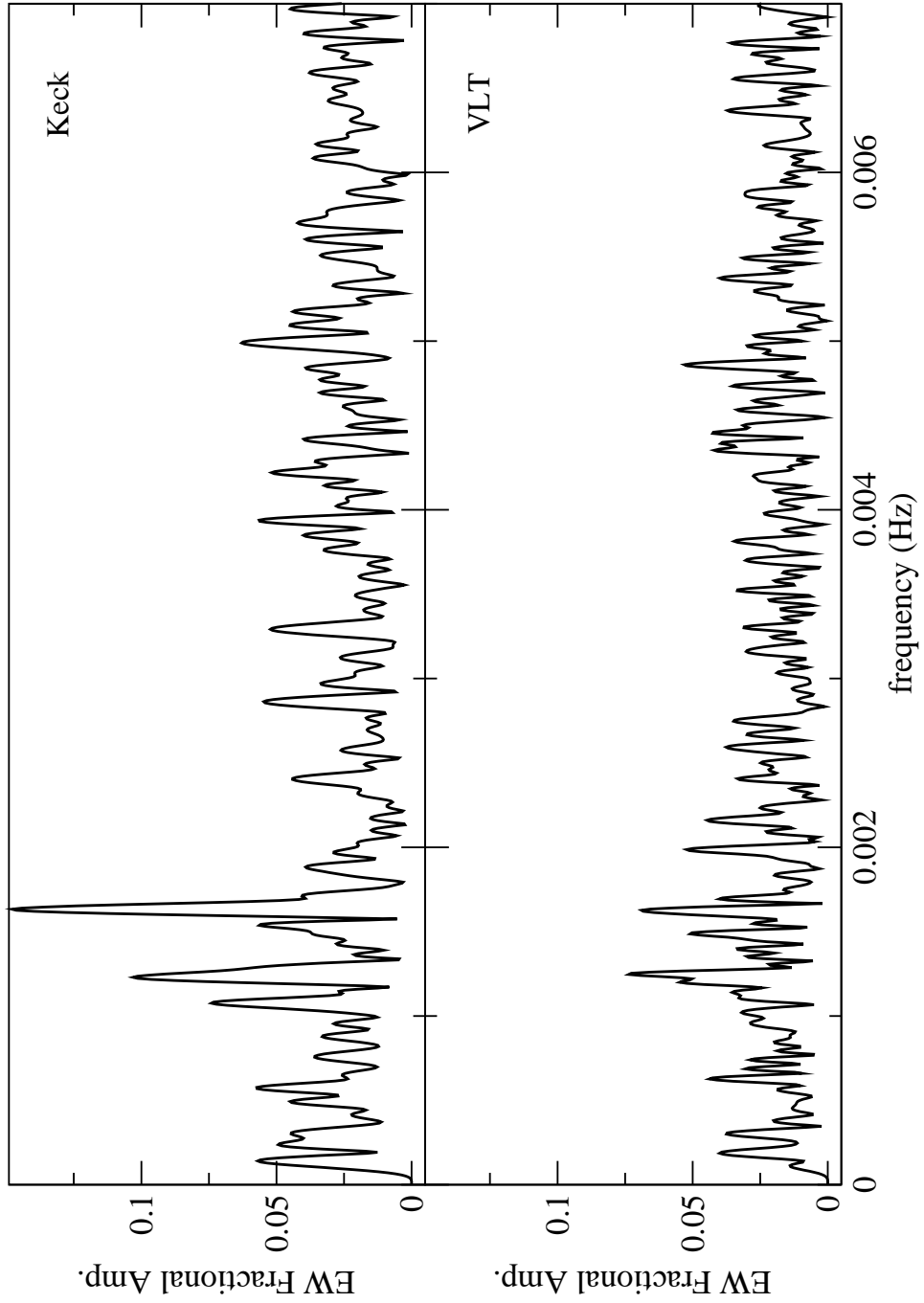} 	
\figcaption{Fourier Transforms of the Ca EWs from the 1996.885 Keck
spectroscopy and the 1999.653 VLT spectroscopy.}
\end{figure}

\begin{figure}[!t]
\plotone{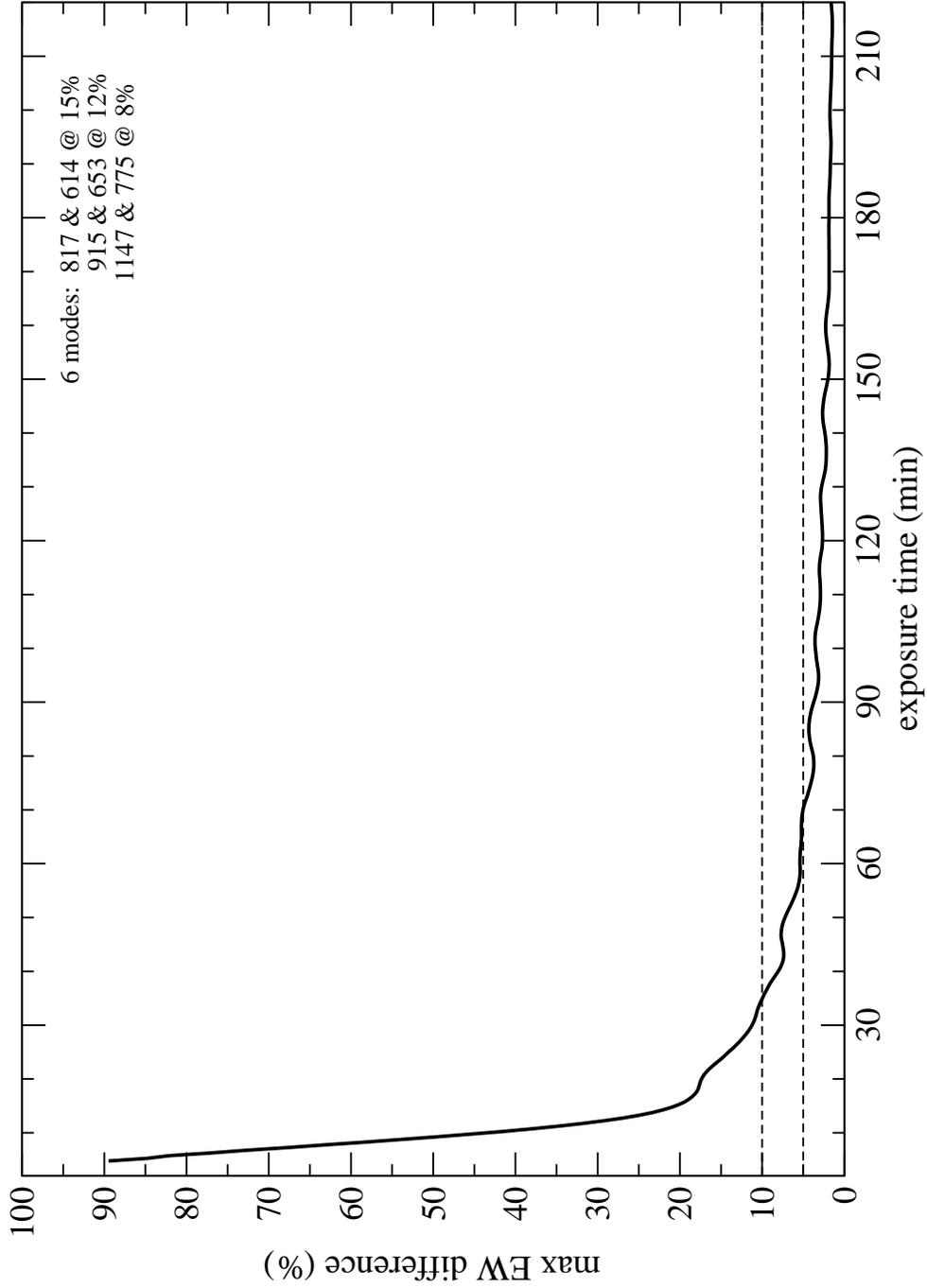}  
\figcaption{The maximum expected contribution of pulsations to EW
variations for G29-38 as a function of the cumulative exposure time.}
\end{figure}

\begin{figure}[!t]
\vspace*{-12mm}
\plotone{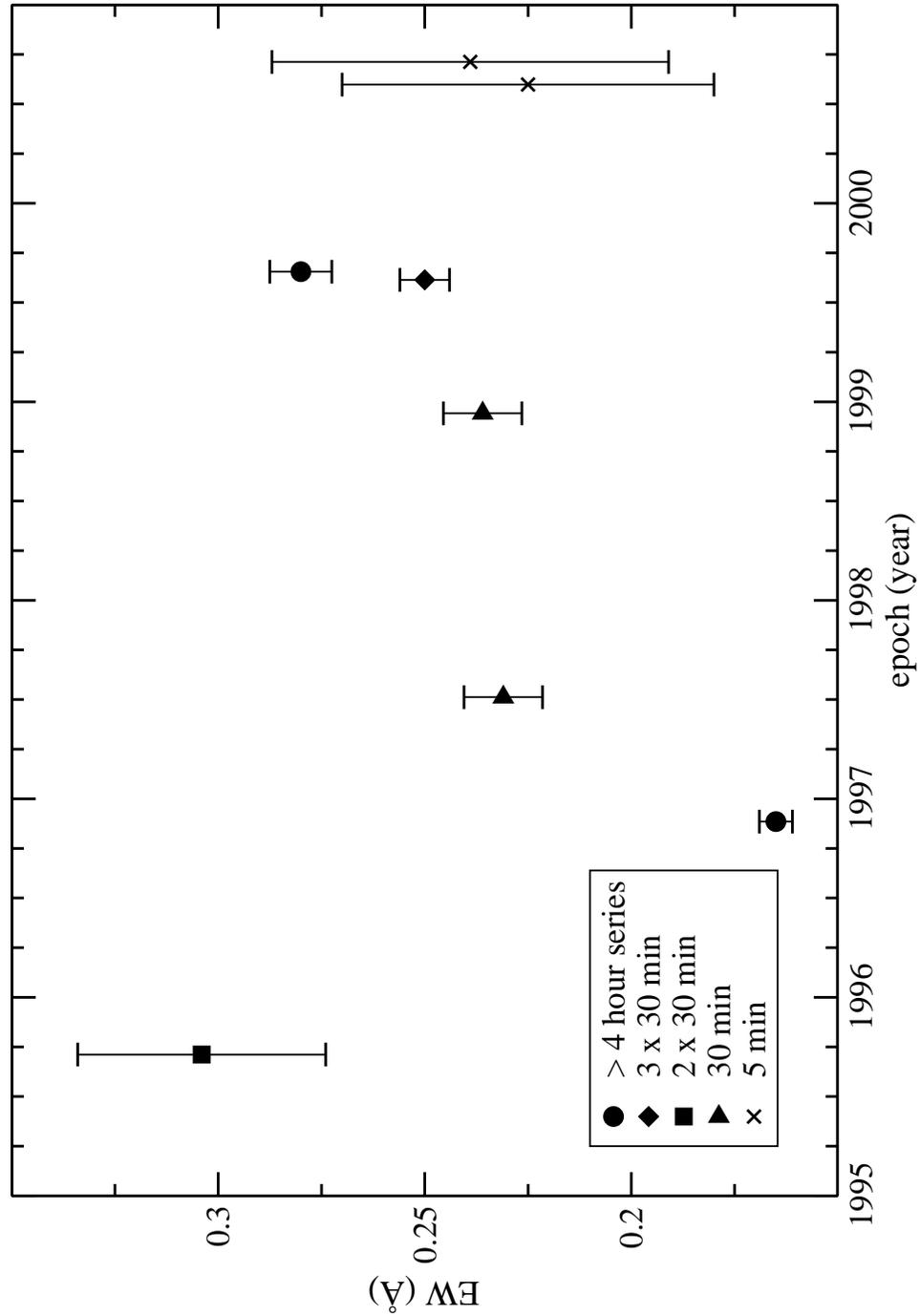}
\figcaption{Calcium EW measurements for G29-38 at eight epochs over a
five year period from three telescopes.  The error bars are a quadrature
addition of the uncertainty in the EW measurement and the expected
variation due to sampling effects.  Both error contributions decrease
with longer exposures.} 
\end{figure}

\end{document}